\documentclass[sigplan,screen,9pt]{acmart}
\renewcommand\footnotetextcopyrightpermission[1]{} 

\acmConference[CoqPL 2018]{International Workshop on Coq for Programming Languages}{January 13, 2018}{Los Angeles, CA, USA}

\setcopyright{none}

\usepackage{booktabs}   
\usepackage{subcaption} 

\usepackage[T1]{fontenc}
\usepackage[utf8]{inputenc}
\usepackage{listings}
\usepackage{todonotes}

\newcommand{\coqatoo}{Coqatoo}

\definecolor{grey}{rgb}{0.8,0.8,0.8}
\definecolor{code-background}{RGB}{255, 248, 220}
\definecolor{code-comment}{RGB}{196, 42, 42}
\definecolor{code-linenumber}{rgb}{0.5,0.5,0.5}
\definecolor{code-keyword}{RGB}{148, 0, 211}
\begin{document}

\title[Coqatoo]{\coqatoo: Generating Natural Language Versions of Coq Proofs}

\author{Andrew Bedford}
\orcid{0000-0003-3101-4272}             
\affiliation{
  \institution{Laval University}            
  \state{Quebec}
  \country{Canada}                    
}
\email{andrew.bedford.1@ulaval.ca}          


\maketitle

\section{Introduction}
Due to their numerous advantages, formal proofs and proof assistants, such as Coq, are becoming increasingly popular. However, one disadvantage of using proof assistants is that the resulting proofs can sometimes be hard to read and understand, particularly for less-experienced users. In an attempt to address this issue, Coscoy et al.~\cite{DBLP:conf/tlca/CoscoyKT95} developed in 1995 an algorithm capable of generating natural language proofs from Coq proof-terms (i.e., calculus of inductive construction $\lambda$-terms) and implemented their approach in two development environments: CtCoq~\cite{CtCoq,bertot1999ctcoq} and its successor Pcoq~\cite{Pcoq,amerkad2001mathematics}. Unfortunately, these development environments are no longer available or maintained; Pcoq's last version dates from 2003 and requires Coq 7.4.

In order to bring this useful feature to modern development environments, we have implemented our own rewriting algorithm: Coqatoo.

\section{Overview of Coqatoo}
Much like Nuprl's text generation algorithm~\cite{DBLP:conf/aaai/Holland-MinkleyBC99}, Coqatoo generates natural language proofs from high-level proof scripts instead of the low-level proof-terms used by Coscoy et al. By doing so, we can avoid the verbosity that comes from using low-level proof-terms~\cite{DBLP:conf/lacl/Coscoy96} and avoid losing valuable information such as the tactics that are used, the user's comments and the variable names. 

Coqatoo's rewriting algorithm can be decomposed in three steps: information extraction, proof tree construction and tactic-based rewriting. 

\paragraph{Step 1: Information extraction}
Using an instance of the \texttt{coqtop} process and the proof script given as input, Coqatoo executes the tactics one by one and captures the intermediary proof states. 

For example, Listing~\ref{listing:before-intros} represents the initial state of Listing~\ref{listing:input}'s proof and Listing~\ref{listing:after-intros} represents the state after executing the first \lstinline{intros} tactic.
\begin{lstlisting}[label=listing:before-intros, captionpos=b,caption={State before executing the first intros tactic}]
  1 subgoal
  
  ============================
  forall P Q R : Prop, (P /\ Q -> R) <-> (P -> Q -> R)
\end{lstlisting}
\begin{lstlisting}[label=listing:after-intros,captionpos=b,caption={State after executing the first intros tactic}]
  1 subgoal
  
  P, Q, R : Prop
  ============================
  (P /\ Q -> R) <-> (P -> Q -> R)
\end{lstlisting}
These intermediary states, which contain the current assumptions and remaining goals, allow us to identify the changes caused by a tactic's execution (e.g., added/removed variables, hypotheses or subgoals).

\begin{figure}[ht]
  \includegraphics[width=0.5\columnwidth]{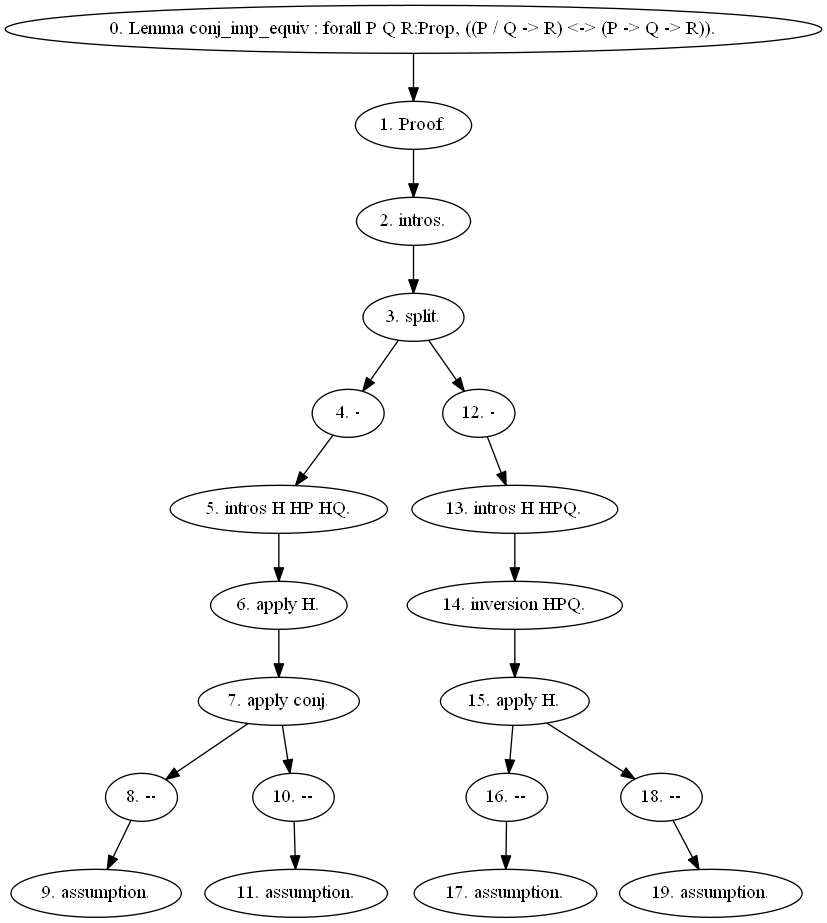}
  \caption{Proof tree of Listing~\ref{listing:input}}
  \label{figure:proof-tree}
\end{figure}

\paragraph{Step 2: Proof tree construction}
We then build a tree representating the proof's structure (e.g., Figure~\ref{figure:proof-tree}). This is a necessary step for our rewriting algorithm as it allows it to determine where bullets should be inserted and when lines should be indented.

\paragraph{Step 3: Tactic-based rewriting}
Finally, we generate the actual final natural language version of the proof using simple rewriting rules. Each supported tactic has its own set of rules. For example, for the \lstinline{intros} tactic we first determine the types of the objects that are introduced. If they are variables, then we produce a sentence of the form \lstinline{"Assume that ... are arbitrary objects of type ..."}. If they are hypotheses, then we instead produce a sentence of the form \lstinline{"Suppose that ... are true"}. Finally, we insert a sentence indicating what is left to prove: \lstinline{"Let us show that ..."}. 

Note that the sentences that we use to produce natural language versions are kept in files that are separate from the code. This allows Coqatoo to support multiple languages and proof styles. For the moment, it can output proofs in English or French, in plain text or in annotation mode (see Listing~\ref{listing:output} for example). In annotation mode, each tactic is accompanied with an informal explanation. We believe that this format will be particularly useful for new Coq users.

\subsection{Example}
To illustrate our approach, consider the proof script in Listing~\ref{listing:input} and Coqatoo's output in Listing~\ref{listing:output}.
\begin{lstlisting}[label=listing:input,captionpos=b,caption=Proof script given as input]
  Lemma conj_imp_equiv : forall P Q R:Prop, 
    ((P /\ Q -> R) <-> (P -> Q -> R)).
  Proof.
    intros. split. intros H HP HQ. apply H. apply conj. assumption. assumption. 
    intros H HPQ. inversion HPQ. apply H. assumption. assumption.
  Qed.
\end{lstlisting}

\begin{figure*}
\begin{lstlisting}[label=listing:output, captionpos=b, caption={Output in annotation mode}]
  Lemma conj_imp_equiv : forall P Q R:Prop, ((P /\ Q -> R) <-> (P -> Q -> R)).
  Proof.
  (* Assume that P, Q and R are arbitrary objects of type Prop. Let us show that (P /\ Q -> R) <-> (P -> Q -> R) is true. *) intros.
  split.
    - (* Case (P /\ Q -> R) -> P -> Q -> R: *) 
      (* Suppose that P, Q and P /\ Q -> R are true. Let us show that R is true. *) intros H HP HQ.
      (* By our hypothesis P /\ Q -> R, we know that R is true if P /\ Q  is true. *) apply H.
      apply conj.
      -- (* Case P: *)
         (* True, because it is one of our assumptions. *) assumption.
      -- (* Case Q: *)
         (* True, because it is one of our assumptions. *) assumption.
    - (* Case (P -> Q -> R) -> P /\ Q -> R: *)
      (* Suppose that P /\ Q and P -> Q -> R are true. Let us show that R is true. *) intros H HPQ.
      (* By inversion on P /\ Q, we know that P, Q are also true. *) inversion HPQ.
      (* By our hypothesis P -> Q -> R, we know that R is true if P and Q are true. *) apply H.
      -- (* Case P: *)
         (* True, because it is one of our assumptions. *) assumption.
      -- (* Case Q: *)
         (* True, because it is one of our assumptions. *) assumption.
  Qed.
\end{lstlisting}
\end{figure*}

\subsection{Comparison}
Compared to Coscoy et al., our approach presents a few disadvantages and advantages.
\paragraph{Disadvantages}
\begin{itemize}
  \item{It only works on proofs whose tactics are supported (see Section~\ref{section:future-work}), while the approach of Coscoy et al. worked on any proof.}
  \item{It may require additional verifications to ensure that unecessary information (e.g., an assertion which isn't used) is not included in the generated proof.}
\end{itemize}

\paragraph{Advantages}
\begin{itemize}
  \item{It enables us to more easily control the size and verbosity of the generated proof (one or two sentences per tactic by default).}
  \item{It maintains the order and structure of the user's original proof script; this is not necessarily the case in Coscoy et al. }
  
\end{itemize}

\section{Future Work}\label{section:future-work}
Coqatoo is only a proof-of-concept for the moment. As such, there remains much to be done before it can be of real use. 

\paragraph{Increase the number of supported tactics}
The number of tactics that it supports is limited to only a handful (see Coqatoo's GitHub repository~\cite{Coqatoo} for more details). We expect that, with the help of the community, we will be able to support enough tactics to generate natural language versions of most proofs in \emph{Software Foundations}~\cite{pierce2010software}.

\paragraph{Add partial support for automation}
In regards to automation, Coqatoo only supports the \lstinline{auto} tactic: if the \lstinline{auto} tactic is present within the script, it is replaced with \lstinline{info_auto} in order to obtain the sequence of tactics that is used by \lstinline{auto}. We plan on adding partial support for automation in the future, starting with the chaining operator "\lstinline{;}". To support this operator we will use our tree representation of proofs to "distribute" tactics on branches.

\paragraph{Integration with development environments} Once it is sufficiently developed, we plan on integrating our utility in modern Coq development environments such as CoqIDE and ProofGeneral.

\paragraph{Add a LaTeX output mode} 
We plan on adding a LaTeX output mode so that the generated proofs can be easily inserted into LaTeX documents.

\begin{acks}
  We would like to thank Josée Desharnais, Nadia Tawbi, Souad El Hatib and the reviewers for their comments.

  We would also like to thank the Coq community for the large number of resources and tutorials that are available online.
\end{acks}

\bibliography{references}

\end{document}